\documentclass{ws-rv9x6}
\usepackage{ws-rv-van,epsfig,graphics}             
\usepackage{ws-index}             
\makeindex
\newindex{aindx}{adx}{and}{Author Index}       
\renewindex{default}{idx}{ind}{Subject Index}  
\begin{document}

\setcounter{chapter}{1}
\chapter{Detection Techniques for Trapped Ions\label{ch2}}

\author[M. Knoop]{Martina Knoop}
\index[aindx]{Knoop, M.} 

\address{CNRS and Aix-Marseille Universit\'e,\\
 Centre de Saint J\'er\^ome, Case C21,
13397 Marseille Cedex 20, France \\
Martina.Knoop@univ-amu.fr}

\begin{abstract}
Various techniques are used to detect the presence of  charged particles stored in electromagnetic traps, their energy, their mass, or their internal states. Detection methods can rely on the variation of the number of trapped particles (destructive methods) or the use of the ion's interaction with electromagnetic radiation as a non-destructive tool to probe the trapped particles. This chapter gives an introduction into various methods, discussing the basic mode of operation completed by the description of recent realizations.

\end{abstract}

\body

\vspace{5mm}The direct observation of a trapped sample can be easily made for macroscopic particles contained in a Paul or Penning trap.  Today, the main interest of the use of ion traps lies in the manipulation and interrogation of atomic or molecular ions. Detection of trapped ions needs amplification and imaging techniques which can rely either on the light emitted by the ions or on their physical presence in the trapping electromagnetic field. Most of today's techniques have been developed not only to make sure that a sample is trapped but moreover to obtain additional information: the size and shape of the ion sample, the number of stored particles, their temperature, the quantum state of the ions, etc.

Two main classes of detection methods exist: destructive techniques \index{destructive detection} which will destroy the trapped ion sample at least partially, and non-destructive techniques, which detect the presence of ions in the trap, keeping their number constant. This latter tool often relies on a perturbing effect to retrieve additional information about the sample. In this chapter, the distinction has been made between techniques which employ the monitoring of photons emitted by the trapped ions (fluorescence techniques), and those which electronically detect the presence of ions in the trap.

The ultimate detection method chosen for a given experiment is of course determined by the species of the trapped ion, the size of the sample, and the dimensions of the trap. Actually, the choice of the detector itself is guided by a large number of technical criteria: the spectral response of the device, its temporal and/or spatial resolution, the gain curve, the dark noise of the device, the environment in which the detector will be installed (vacuum conditions, contaminants, ..), as well as the electric and magnetic environment. This chapter gives a review of the major techniques and instruments used, without being exhaustive. In every paragraph, additional references will give details on the described technique or present major results of its implementation.

\section{Electronic Techniques}\label{sec:dest_tec}
Many trapped ions cannot be made visible by resonant excitation, either because they do not dispose of a closed fluorescence cycle giving a large enough number of scattered photons, or because the wavelength corresponding to the absorption and emission of photons is difficult to generate with a sufficiently powerful source. Due to experimental and geometric constraints, it is much easier to obtain a reasonable signal-to-noise ratio of the sample of trapped ions if the excitation process is performed by a laser rather than a lamp. However, the generation of a resonant laser wavelength is technically more demanding than its non-coherent counterpart.

Over the years, different methods have been developed in order to detect the presence of ions in the trap, even if they do not emit fluorescence. After a review of the different detectors employed, the main techniques for electronic detection of the trapped sample will be presented.

\subsection{Instruments}\label{ssec:instr}
As mentioned in the introduction, technical specifications, such as the spectral response or the temporal resolution, are decisive in the selection of the best detector. Three main devices exist and are in use today. \cite{knoll10}

\subsubsection{Faraday Cup}
The Faraday cup is a metallic recipient open on one side in order to absorb a beam of charged particles. The cup is grounded via an electrometer capable of measuring currents in the pico Ampere range. The detection limit is of course due to the sensitivity of the electrometer. The device is primarily destined to measure large particle numbers and can reach a lower detection threshold of a couple of hundred ions in the best case scenario.

\subsubsection{Electron multiplier}\index{electron multiplier}
Electron multipliers (EMs) are versatile tools for ion or electron detection. They are composed of a large number of stacked polarized dynodes which will produce an amplified cascade of secondary electrons. In general, the first dynode is polarized around -3 k$eV$ for positive ion detection. It is possible to detect single ions with this device as the gain can reach 10$^8$ and the dark count under high vacuum conditions is below one count per second. The effective detector surface is very similar to that of a photomultiplier, typically around 100 mm$^2$.
EMs exist in two versions: the discrete-dynode and the continuous-dynode (or channel) EM. EMs must work under reduced pressure conditions.  Due to the electronic conversion of analog pulse signals into transistor-transistor logic (TTL)-pulses, an EM always presents a dead time in its operation process. Durations below 10~ns have been achieved \cite{liu93}.
EMs are often coupled to a pre-amplifier and a discriminator generating a direct TTL output.

\subsubsection{Microchannel Plate}
\begin{figure}[h!]
\centerline{\psfig{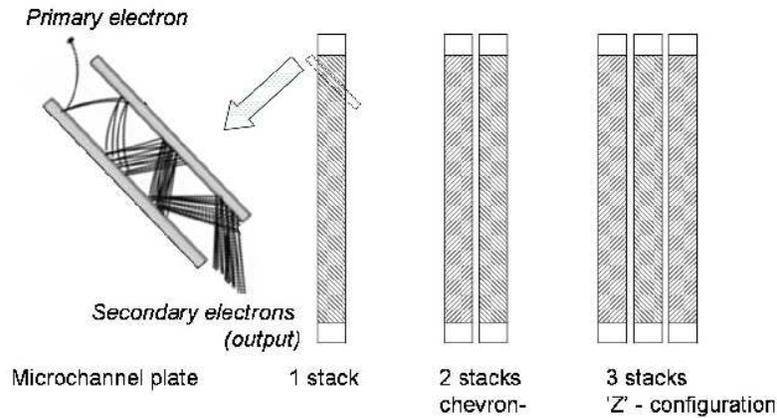}}
\caption{Scheme of an MCP. Left: the principle of operation, right: stacking multiple MCPs increases the gain of the detector. }
\label{fig:mcp}
\end{figure}

\index{microchannel plate}
Microchannel plates (MCPs) rely on the same principle of operation as EMs: a cascade of secondary electrons is generated in an array of very small continuous-dynode amplifiers
\cite{wiza79}. The device is composed of a million channel-EMs each having a diameter of about 10~$\mu$m. The gain of the MCP is proportional to  the length to diameter ratio, $L/d$, of the individual channels following the relation $g = \exp(G \times(L/d))$, where $G$ is the secondary emission characteristic of the channel called gain factor. $L/d$ can be as high as several hundred, for a total absolute gain value above 10$^4$. Higher gain is achieved by MCPs configured  with multiple sets of channels mounted in line. Depending on their respective geometry this can be in the ``chevron" mode (two stacks), or in the three-stack ``Z" configuration, see Figure \ref{fig:mcp}.

An MCP  has an entry converter in order to convert the impinging particles into photons or electrons. Furthermore, the MCP stack  is always followed by a read-out device to detect the produced electron avalanche. Depending on the application this can be a simple anode giving a one-dimensional output (or a more complex anode array for a two-dimensional signal), or a phosphor screen followed by a charged-coupled device (CCD) in order to keep the spatial information of the detector.

The temporal resolution of the MCP is limited by the electronics and read-out device, but is typically inferior to a nanosecond.

\subsection{Techniques}\label{ssec:tech}
Many of the electronic detection techniques rely on the ejection of the trapped ion cloud in order to be monitored by a detector external to the storage device.\index{destructive detection} However, as the contained sample is always charged, the image current which the sample induces on the electrodes is a non-destructive means for detection. A short description of some techniques is given in Werth's book on charged particle traps \cite{major05}.

\subsubsection{Ion loss}
The most simple detection method of a trapped ion sample is to switch off the trapping field and record a synchronized signal on an ion detector. The fine-tuning of this measurement will have to take into account the solid angle under which the ions leave the trapping region, their flight path to the detector, and  the phase of the trapping field at the moment it is turned off.

More sophisticated versions of detection will therefore make use of  ion optics in order to collect as many charged particles as possible and guide them to the detection region. As ``turning off" the trapping field is often not an instantaneous action (in particular if a magnetic field or a radiofrequency field is present), a technically simpler solution is the ejection of the ions with an acceleration pulse without modification of the trapping field.
Ions can be ejected along all axes of the trapping field.

A first application of this technique is to measure the trapping time constant of the trap. Recording the relative ion signal for different durations since the first filling of the trap gives the loss rate of the trap. Due to space charge effects, it can be shown that a completely filled trap will lose ions more rapidly than a trap which contains only a small ion cloud. The observed exponential decay of the signal is often ruled by two time constants: one which is due to the trap and the depth of its potential, and one which is determined by the collisions with the residual gas of the vacuum vessel. This latter influence can also be beneficial for the ion cloud, as the addition of a light atomic gas in the vacuum vessel can cool the ion cloud as a consequence of elastic collisions.

\subsubsection{Depletion techniques}\index{depletion techniques}
These  detection methods record the variation of the trapped ion number as a function of additional interaction processes. The technique is often employed in chemistry or physical chemistry experiments in order to investigate the reaction of a trapped species with an introduced reactant gas or an applied source of electromagnetic radiation. The variation of the number of trapped ions gives an insight into the rate of the process, under the condition that pressure and temperature parameters are extremely well controlled.

In order to \emph{actively }produce a spectrum, the main idea is to modify the $m/Z$ ratio of the trapped molecular species by absorption of one or several photons. Depending on the energy of the photon, various reactions can be triggered. Photofragmentation of a chemical bond requires an ultraviolet (UV) photon or multiple infrared (IR) photons, whereas photodissociation can be made by one high-frequency IR or multiple IR photons. Even lower energies are required for the photodissocaition of a weak bond or the photodetachment of an electron. For a detailed review see Rizzo \textit{et al.}\cite{rizzo09}

A recent example is the monitoring of the resonance enhanced multiphoton dissociation in $H_{2}^{+}$-ions, where the study of the photodissociation process has been made by monitoring the number of ions still present in the trap as a function of the number of laser shots and of the energy of the dissociation laser \cite{karr12}.

\subsubsection{Ejection with or without additional perturbation}\index{ion ejection}
Active ejection of the ions from the trap is a way to measure relative ion numbers. As mentioned above, an optimized experiment will take into account the phase of the radiofrequency trapping field for the timing of the ejection pulse. Higher sensitivity in the detection rate can be achieved by collecting and guiding the ejected ions by a set of (electrostatic) Einzel lenses. Optimization of experimental features can be made numerically with good success \cite{simion}.

A major experiment in ion trapping using this technique is the work carried out in the Mainz group on the trapping performances of a radiofrequency trap. \index{nonlinear resonances}\index{resonance} Alheit \textit{et al. }\cite{alheit95} recorded the number of trapped ions for the complete first stability range using samples of a few thousand $H_2^+$-ions in a trap with a radius of 20~mm. With this experiment, they were able to characterize the ion trap and demonstrate that the trapping behavior in a large interval of storage parameters is not homogeneous. Actually, at a given resonance condition,
\begin{equation}
n_r \omega_r + n_z \omega_z = \kappa \Omega ,
\end{equation}
with $\kappa$ an integer value, and $|n_r|+|n_z|=N$, resonances of order $N$ appear in the trapping zone. If this condition is fulfilled, ions can take up energy from the trapping field, they are heated and then lost from the trap. Strong absences in the trapping capacity appear if these nonlinear resonance conditions are fulfilled. The phenomenon is, of course, amplified for larger clouds, as space charge becomes an issue for all ion numbers larger than one.
The same feature can be used for isotope separation \cite{alheit96b}, as the trap operation is different for varying $m/Z$-ratios. \index{ion heating}

\begin{figure}
\centerline{\psfig{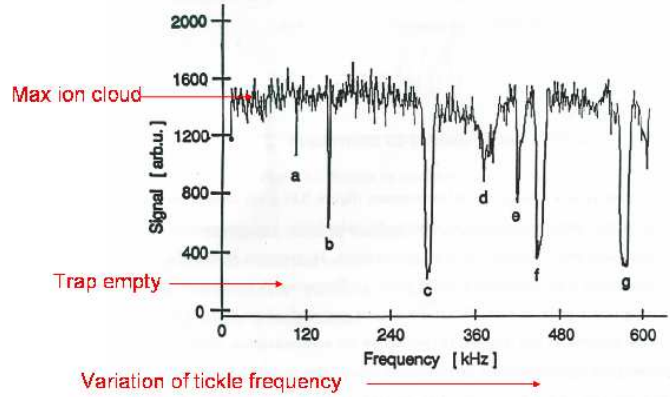}}
\caption{Motional frequency spectrum of a trapped cloud of calcium ions at trapping frequency $\Omega/2\pi = 1$~MHz. Different frequencies and their coupling terms can be identified : $\mathbf{a}: \omega_r/2$, $\mathbf{b}:  \omega_r$, $\mathbf{c}:  \omega_z$, $\mathbf{d}:  \omega_z + \omega_r/2$, $\mathbf{e}: 2\omega_z -\omega_r$, $\mathbf{f}: \omega_r + \omega_z$, $\mathbf{g}: 2\omega_z$. }
\label{fig:ms}
\end{figure}

Rather than scanning the whole stability range, insight into the actual trapping parameters can also be gained by using resonant excitation of the stored ions. The addition of a small oscillating voltage on the trap electrodes can excite motional frequencies if the frequency of the oscillating voltage coincides with the a macromotion component of the cloud. It is thus possible to record the spectrum of secular frequencies $\omega_i$, by repeating a protocol of application of the described additional ``tickle" frequency, and a subsequent ejection of the ion cloud, measuring \textit{de facto} the residual ion number as a function of the applied excitation frequency \cite{vedel90b} (see Figure \ref{fig:ms}).\index{frequency spectrum}

\subsubsection{Time-of-flight profiles} \index{time-of-flight method}
An even more refined technique of the ejection method is the recording of the time-of-flight (TOF) profile of an ejected ion cloud on a detector situated at a fixed distance. The distribution of arrival  times of the ejected ions  will reflect their mass distribution, and it can be described by a set of experimental parameters as there are the $m/Z$ ratio of the species, the velocity distribution of the ions in the trap, and the distance, $d$, of the detector.
The longer the flight path  the better the temporal resolution of ions arriving at the detector. A typical experiment is composed of an extraction zone next to the trap, followed by a drift zone before the detector. An energy ramp might eventually be added in order to modify the potential energy of the ions with respect to the detector \cite{lunney92}. For a single species in a Paul trap, the recorded profiles of an ejected ion cloud show a double bump as a result of the velocity distribution of the ions in the trap. Moreover radiofrequency residuals might also distort the profile.
The extension of the flight path by means of ion mirrors to almost infinite length in a multi-turn device allows the increase of the mass resolution $m/\Delta m$ beyond 30 000 \cite{shimma10}.

An extension to the time-of-flight method coupled with the excitation of an ion cyclotron resonance (ICR) makes it possible to reach unprecedented precision in the mass resolution \cite{blaum06}. The precise determination of the cyclotron frequency $\omega_c = q B/(2 \pi m)$  of a single cooled ion is the basis for this advanced technique of mass measurement in Penning traps.

The ion with negligible radial kinetic energy is excited to a well-defined magnetron radius by external dipole excitation. Application of an additional oscillating quadrupole field on the segmented trap converts the magnetron into cyclotron motion. This manipulation drastically increases the radial energy, as the frequency values for the various motional states differ by several orders of magnitude. Once the ion is  ejected from the trap,  its radial energy is transformed into axial energy by coupling the magnetic moment of the orbital motion to a magnetic field gradient. It is then possible to record the time-of-flight of the ejected ions as a function of the exciting frequency, the shortest time-of-flight value corresponding to the cyclotron frequency of the ions. This technique has been realized with mass resolving powers of more than $10^7$.

Even higher resolutions can be achieved by applying the exciting pulse not in a continuous way, but using a temporal Ramsey method to achieve still narrower resonance lines. Like many other methods, absolute values can only be determined by calibration with a well-known reference mass. \index{Ramsey method}
Best results (i.e., a good signal-to-noise ratio) are obtained with a single ion in the trap and a few hundred repetitions.

\subsubsection{Image currents}\index{image current}

The recording of the image current which is induced by a trapped ion cloud onto the trap's electrodes allows non-destructive ion detection \cite{wineland75a}. The oscillation of the trapped ions leads to an oscillating current 
between the trap electrodes. 
The sensitivity of this method can reach the single-ion level. The signal-to-noise ratio for this detection method is given by
\begin{equation}
 S/N = \frac{\sqrt{\pi}}{2} \frac{r_{ion}}{D} Z \sqrt{\frac{\nu}{\Delta\nu}} \sqrt{\frac{Q}{kTC}} ,
 \end{equation}
with $r_{ion}$ the ion's motional radius and $Z$ its charge. $D$ is the effective electrode distance, and $T$,  $C$, and $Q$ the temperature, capacity, and quality factor of the detection system, respectively. $\frac{\nu}{\Delta\nu}$ is the ratio of the ion's motional frequency to its width.
In the case of an individual singly charged ion, the induced image current is only about a few hundred femtoampere (fA). For multiply charged ions, this current increases proportionally to the charge. Thermal noise is one of the main limiting factors in these experiments, and better results can be obtained by cooling the detection electronics and the traps either with liquid nitrogen or with liquid helium.

Detection and resistive cooling of a single trapped ion was demonstrated as early as 1998 in a cryogenic Penning trap \cite{diederich98}. \index{resistive cooling} In this experiment, the narrow-band detection circuit  consisted of a superconducting coil which formed a parallel LC resonant circuit with the parasitic capacities of the device. The signal-to-noise ratio was determined by the quality factor $Q$ of the circuit, which had to reach a value of a few thousand \cite{haeffner03}. The image current was detected by the proportional voltage drop across the trap, which was amplified, and then Fourier transformed in order to determine the corresponding frequency spectrum. \index{frequency spectrum} Necessary calibrations were made by trapping reference ions in the setup.

\section{Fluorescence Techniques}\label{sec:nondes_tec}
The interaction of atoms with light is one of the most powerful tools developed in ion-trapping experiments. Laser light is often used to create the ions, to prepare them in different quantum states, to cool them, or to heat them. Using the photons emitted upon (resonant) excitation of an atomic transition is the most reliable way to detect and identify the trapped particles. \index{ion fluorescence}

\begin{figure}[h!]
\centerline{\psfig{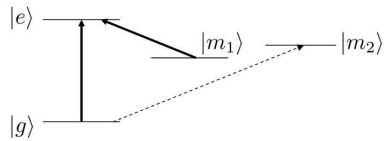}}
\caption{First energy levels of a typical three-level ion. $|g\rangle$, $|e\rangle$, and $|m\rangle$ denote the ground, excited, and metastable states, respectively. }
\label{fig:en_schem}
\end{figure}

The direct detection of the ion's fluorescence is typically made on a strong transition, most often the resonance line.  Many ions trapped today have energy schemes very similar to that pictured in Figure \ref{fig:en_schem}. The strongest transition (the resonance line) is the one connecting the ground state $|g\rangle$ to the excited state $|e\rangle$. The lifetime of the excited state $|e\rangle$ is of the order of some tens of ns, giving rise to a natural linewidth of the corresponding atomic transition of the order of some tens of MHz. The metastable states $|m_i\rangle$ have a lifetime of the order of a second, the transition connecting them to the ground state are forbidden in the dipole approximation. The branching ratio from the $|e\rangle$-state towards the metastable states is typically several percent compared to the de-excitation towards the ground state. Given the large difference in lifetimes, a repumper laser from $|m_1\rangle$ to $|e\rangle$ might be necessary in order to close the cycle and  obtain an observable fluorescence signal. Quenching with a buffer gas is a less efficient but simple alternative method; inelastic collisions with a heavy or molecular buffer gas reduce the lifetime of the metastable state \cite{knoop94}. \index{metastable state}

If the resonance line is excited by a laser reaching the saturation limit, a few million photons can be scattered per second per atom. Taking into account the finite collection angle, the losses in the optical system, and the quantum efficiency of the detector, a typical setup has a detection efficiency of about 10$^{-4}$. A single-ion signal is then about 10 000 photons per second and can easily be recorded with a photomultiplier in photon-counting mode. A good imaging optics is, however, mandatory in order to spatially separate the ion's fluorescence signal from the background noise due to scattered light by the exciting laser beams (see Section \ref{sec:optsetup} for details).\index{detection efficiency}

The most direct example of fluorescence detection is the observation of a single Ba$^+$-ion by the human eye, performed in an experiment at the University of Hamburg \cite{neuhauser78b}.
In fact, this experiment had a microscope objective mounted on one of its viewports, and the trapped ion could be directly visualized during its life in the trap. The necessary condition for this experiment is, of course, to have a strong transition in the visible, preferably in the center of the eye's spectral sensitivity.

Absorption imaging is a standard tool in ultracold matter experiments  with neutral atoms; it is not commonly used in ion-trapping experiments due to the very low density of ion clouds. Only very recently, a group at Griffith University has recorded the absorption image of a single Yb$^+$-ion \cite{streed12}. The observed image contrast of 3.1(3)\% is the maximum theoretically allowed for the imaging resolution of the employed  setup.

\subsection{Lineshape}
The observation of the fluorescence of an atomic ensemble can also be made while varying the frequency of the exciting laser beam. The spectral response of the ions will reflect the excitation probability as a function of wavelength and will provide a multitude of additional information. The recorded linewidth is, in most cases, larger than the theoretical natural linewidth of the transition. The main broadening factor is the Doppler effect, which can generate linewidths of the order of several GHz for an uncooled ion cloud. Many experiments use laser-cooling techniques, which make the recording of a Doppler profile slightly more complex, as probing and cooling are not uncorrelated. A good estimation of the temperature of a laser-cooled target is a profile of the repumping transition keeping the laser on the resonance line fixed.

Scanning the cooling laser frequency may provoke a structural transition from a cloud to a Coulomb crystal \index{Coulomb crystal} configuration where the ions are better localized. Structural phase transitions of the atomic ensemble can be indicated by a change in the line profile, as they are marked by a change in the recorded atomic lineshape \cite{diedrich87}.

Under the influence of laser cooling and in a trap where micromotion perturbations are compensated, the ion may reach the strong coupling limit where its motional amplitude is smaller than a fraction of the excited wavelength \cite{dicke53}. The spectral response of the ion will then split up into a central carrier and sidebands, which can serve to determine the vibrational state that the ions are in \cite{wineland79,diedrich89}. For an ion cooled to the lowest vibrational levels, the number of sidebands is reduced to very few, the ion is said to be in the Lamb--Dicke regime.\index{Lamb--Dicke regime}

The monitored lineshape can also give information on the environment of the trapped ions. It can be used to measure the influence (or absence) of collisions and of magnetic or electric fields. Depending on the geometric configuration of laser beams, dark resonances \index{dark resonance} appear in a three-level system when a coherent superposition of the involved states is created. These dark states depend on the involved laser linewidths and powers, but also on the micromotion amplitude of the probed ions \cite{lisowski05b}.

\subsection{Single-ion detection}\index{single-ion detection}
Counting the number of trapped ions is a perfect way to control the experimental conditions, to calibrate the necessary parameters, and to obtain a complete insight into the underlying physics process. However, the determination of a precise ion number can only be made after having identified the net fluorescence signal for a single ion. Quantum jumps can be observed in a three-level atom, when one of the excited states is a metastable state (Figure \ref{fig:en_schem}). In this configuration, the cooling laser beams are applied on the cycling transitions $|g\rangle - |e\rangle - |m_1\rangle$, and the fluorescence is observed on the resonance line. Applying an additional excitation to the $|g\rangle - |m_2\rangle$ transition will provide optical pumping to the $|m_2\rangle$- state which has a very long lifetime \cite{dehmelt75a}. With a single trapped ion, the observed fluorescence signal will then become binary: ``ON"  when the ion is scattering photons on the cooling cycle transitions, and ``OFF" when the ion is pumped to the metastable state \cite{nagourney86,sauter86,bergquist86}. The duration that the ion will remain in the ``shelved" dark state corresponds  to the lifetime of the level \cite{knoop04}. \index{quantum jumps}

A single trapped ion is a well-localized point source; storing two ions in a trap allows the investigation of the interaction between two extremely well-defined sources. A spectacular realization of this is Young's double-slit experiment \index{Young's double-slit experiment} which recorded the interferences of two laser-cooled ions trapped within the same device.
\cite{eichmann93}

\subsection{Motional frequencies}

\begin{figure}[h!]
\centerline{\psfig{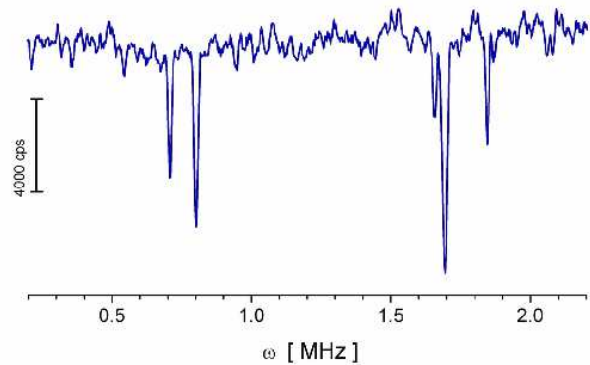}}
\caption{Motional frequency spectrum of a laser-cooled cloud of calcium ions. On this graph the ion's fluorescence is recorded while varying the excitation tickle frequency. }
\label{fig:ms2}
\end{figure}

\paragraph{Secular motion} Measuring the secular motional frequencies (macromotion) of the trapped ions provides information about the stiffness of the trap, eventual screening effects, and all possible asymmetries. Many electronic methods  have been developed in order to precisely measure frequencies since this is the basic tool for all mass spectrometry experiments (see Section \ref{sec:dest_tec}).
Motional frequencies can also be measured via the fluorescence emitted by the ions, the easiest application using a cloud of laser-cooled ions. \index{frequency spectrum} If an additional small oscillating voltage (a ``tickle") is applied to the trap, the ions will absorb energy and be slightly heated, if they are in resonance with the tickle frequency. This response can be made visible on the fluorescence signal by a drop in the maximum amplitude (see Figure \ref{fig:ms2}). Scanning the tickle frequency will thus reveal all motional frequencies of the trapped ions if the applied tickle amplitude is not too strong and the scanning frequency not too fast, as the cloud has to be re-cooled after each excitation \cite{champenois01}.

\vspace{1mm}
\paragraph{Micromotion}\index{micromotion} The micromotion amplitude of a trapped ion increases with its distance from the trap center. Moving ions in the trap is a way of obtaining information about the trapping field. Manipulation of the probe can be made by applying additional static voltages. All traps which are destined to trap a single ion are, in general, laid out to reach the Lamb--Dicke regime. In order to reach this, and to correct for eventual asymmetries of the trap (due to mounting or coating by the atomic beam during ion creation), these traps are mounted with a set of compensation electrodes, which allow -- via the application of direct current (DC) voltages -- the movement of the ion in the trap potential, such that it reaches the minimum of the trapping field.

Different techniques are used in order to compensate for excess micromotion which is induced by a defect or asymmetry in the potential. Large asymmetries can be observed on a camera, as the ion (or ion cloud) will move if the potential well is lowered. The fine tuning of the micromotion compensation is often carried out by recording  the correlation signal of the photon count with the trapping radiofrequency \cite{berkeland98}. The more micromotion the ions experience the larger the amplitude of this correlation signal, and this therefore constitutes an extremely precise tool for canceling all defects of the trapping potential.

\subsection{Ions as a spatial probe}
In a seminal experiment, Guth\"{o}hrlein \textit{et al.} used a trapped ion to explore the mode structure of the cavity which has been superimposed onto the trap \cite{guthoehrlein01}. They  probed different Hermite--Gauss modes with an axial resolution of 170~nm, where the excitation rate of the trapped ion served as a sensitive detector.

\subsection{Temporal Ramsey}
The first  atomic frequency standard based on the interrogation of laser-cooled ions  was reported in 1985 by the National Institute of Standards and Technology (NIST) Ion Storage Group \cite{bollinger85}. The probed clock transition was a hyperfine transition of the Be$^+$-ion in the microwave region. Ramsey's method of separated oscillatory fields has been implemented in a temporal regime. Actually, a radiofrequency pulse of duration $t$ was applied, followed by a free precession interval of variable duration $T$ (which could be as large as 19~s) and a second radiofrequency pulse of duration $t$ coherent with the first pulse. This technique permitted the resolution of the atomic transition to better than 25~mHz, corresponding to a residual systematic uncertainty of $9.4 \times 10^{-14}$.

\subsection{Quantum logic spectroscopy} \index{quantum logic spectroscopy}
Recently, the development of an innovative probing method has allowed the application of precision spectroscopy to ions that do not have a suitable atomic transition for laser cooling or atomic state preparation \cite{schmidt05}. In a quantum logic spectroscopy experiment, two ions of different species are trapped. The first, ``logic", ion is laser cooled, its atomic states are prepared via optical pumping and the read-out is made via the recording of its fluorescence. This ion provides sympathetic cooling to the second, ``spectroscopy", ion. Entanglement of both  ions is made via the common motional modes, and this allows  a coherent transfer of the
spectroscopy ion's internal state onto the logic ion, where it can then be measured with high efficiency. This technique allows the probing of ions which are not directly ``visible" in the experiment, and is the basis for the most accurate optical atomic clocks \cite{chou10}.

\subsection{Imaging techniques}
The  visual observation of a single Ba$^+$-ion by Toschek's group in Heidelberg \cite{neuhauser80} was the first in a long line of spatial resolution images that have been taken of trapped ions. A couple of years later the first images of crystallized structures were published by Walther's group in Munich \cite{diedrich89}. Today, images of ion clouds and crystals with single-ion resolution are a standard tool in Paul trap experiments \cite{hornekaer02}. The aspect ratio of a laser-cooled cloud gives insight into the stiffness of axial and radial potential in a linear trap. Dark sites in the crystal (due to other isotopes or other ions), allow the monitoring of the rearrangement of the ions on the crystal sites. Reaction rates of trapped ions can be determined by recording  cloud sizes, shapes, and evolutions in the controlled atmosphere of a reactant \cite{willitsch08,hall11}.
In a perfect potential the observed crystal structures should rotate around the symmetry axis. Simulations have shown that very small defects in the potential introduce sufficient asymmetry for this not to happen in Paul traps \cite{marciante10}. In Penning traps, however, laser cooling will generate a rotating plasma of stable structure. Active stroboscopic techniques have to be implemented in order to synchronize the observation to the plasma rotating frequency \cite{huang98}. This latter technique has allowed the direct observation of the structural phases of a crystallized ion plasma \cite{mitchell98}.

\subsection{Optical setup and instrumentation}
\label{sec:optsetup}
Different detectors can be used in order to record the fluorescence of the trapped atoms. Traditionally, a photomultiplier (most often in the photon-counting mode) will collect the emitted photons. In order to gain information about the spatial distribution, CCD cameras are often used. These devices have to be coupled to  image intensifiers in order to amplify the signal emitted by individual ions. Actually, intensifiers consist of a (wavelength sensitive) photocathode, an MCP, and a phosphor screen; these can be integrated to the camera or purchased externally. One drawback of the use of a camera is the slow response rate, rarely more than about a hundred images per second. Gateable devices can be synchronized onto well-defined trigger events and can overcome this limitation. Electron multiplier CCDs (EMCCDs) add supplementary gain to the detection process by impact ionization, although they are not gateable.

For the design of the optical setup, various external parameters have to be taken into account. The numerical aperture of the imaging objective will determine the solid angle for photon collection, which is one of the main factors in the determination of the detection efficiency. The working distance of the objective will depend on the position of the objective. If it is inside the vacuum vessel, it can be placed close to the trap, a perturbation of the trapping potential by the  isolating (glass) surfaces must however be avoided. A recent idea \cite{lange09pr} is the collection of atomic fluorescence by an optical fiber which is easy to implement in the trapping device.

In any configuration, the objective will be aligned in order to give an intermediate image. Only in this image plane can spatial filtering be made without losses of the signal amplitude; a diaphragm or pinhole can separate the atomic fluorescence signal from the scattered light of the exciting laser, which is in many cases at the same wavelength. The choice of the optics is vast, it can reach from multi-surface systems to very simple lenses \cite{berkeland02}, depending on the ultimate goal of the experiment. A very clever use of the individual ion detection is made by the Innsbruck group, where the high-resolution optical setup also serves to individually address single ions in a trapped chain \cite{naegerl99}.

\bibliographystyle{ws-rv-van}

\printindex                         
\end{document}